\documentclass[12pt]{article}

\usepackage{epsf}
\usepackage[dvips]{graphicx}
\newcommand{\CA}{{\cal A}}

\newcommand{\CD}{{\cal D}}

\begin{document}
\def\bp{{\mbox{\boldmath$p$}}}
\def\bP{{\mbox{\boldmath$P$}}}
\def\bk{{\mbox{\boldmath$k$}}}
\def\br{{\mbox{\boldmath$r$}}}
\def\bq{{\mbox{\boldmath$q$}}}
\def\bn{{\mbox{\boldmath$n$}}}
\def\ba{{\mbox{\boldmath$a$}}}
\def\bb{{\mbox{\boldmath$b$}}}
\def\bc{{\mbox{\boldmath$c$}}}
\def\bxi{{\mbox{\boldmath$\xi$}}}
\def\bsigma{{\mbox{\boldmath$\sigma$}}}

\begin{center}
{\large \bf CHARGE-EXCHANGE REACTION
\boldmath{$pD \to n(pp)$}
WITHIN THE BETHE-SALPETER APPROACH}

\vskip 5mm

S.S. Semikh$^{1}$, S.M. Dorkin$^{2}$, L.P. Kaptari$^{1}$ and B.
K\"ampfer$^{3}$

\vskip 5mm

{\small
(1) {\it
BLTP, Joint Institute for Nuclear Research, Dubna, Russia
}
\\
(2) {\it
Far Eastern State University, Vladivostok, Russia
}
\\
(3) {\it
Forschungszentrum Rossendorf, Germany
}
}
\end{center}

\vskip 5mm

\begin{center}
\begin{minipage}{150mm}
\centerline{\bf Abstract}
The deuteron charge-exchange reaction $pD\to n(pp)$, for low
values of the momentum transfer and small excitation energies
of the final pp pair, is considered within the framework of the
Bethe-Salpeter approach. A
method for calculating observables is developed for the case
of the pp pair in a $^1S_0$ state. Results of
numerical calculations for the plane-wave approximation are
presented.
\\
{\bf Key-words:}
Bethe-Salpeter formalism, deuteron, charge-exchange,
deuteron tensor polarimeter.
\end{minipage}
\end{center}

\vskip 10mm

\section{Introduction}

We present here the theoretical framework for the
charge-exchange reaction
$pD \to n(pp)$ within the Bethe-Salpeter (BS) approach.
This reaction is a well-known process, and its investigation
was started more than 20 years ago
in Dubna and Saclay, in particular at the famous SATURN-II setup.
This process was experimentally investigated in detail in Saclay
for kinetic
energies of the incoming deuteron of 200 and 350 MeV.
Data of the cross section and two tensor analyzing powers for different
kinematical regions are available \cite{kox}.
Let us briefly recall why it was so interesting
to explore this process. At the very beginning there were
some experimental indications and
strong theoretical evidences that this reaction
has a large cross section and, at the same time, the tensor analyzing
powers are also significant up to deuteron beam energies of 500 MeV.
So it was immediately
suggested to build a deuteron tensor polarimeter exploiting this
reaction. Indeed, such a polarimeter was successfully constructed
\cite{polder}. Recently, a renewed interest
in such a device has arisen and,
therefore, a theoretical re-examination of the reaction $pD\to n(pp)$
is desired. We are going to consider this process for higher
proton beam
energies available at the cooler synchrotron COSY in J\"ulich.
The main goal of our
investigation here is to check whether the deuteron tensor
polarimeter seems feasible for the COSY energies, too. But first of
all we have to test our model for the low beam energies by comparing
with the experimental data obtained at SATURN-II. In our consideration we
follow the basic ideas in Refs. \cite{wil1}--\cite{wil2}.

\section{Calculations}

The process $pD\to n(pp)$ is considered in the deuteron rest
frame where the deuteron has the 4-momentum
$p_D = (M_D, {\bf 0})$. In the initial state there is a fast proton
with 4-momentum $p_p = (E_p, \bp)$.
The final neutron with 4-momentum $p_n = (E_n, \bn)$ is assumed to
have almost the same momentum as the initial proton. This means that we
consider only small values of the transferred momentum described by
the Mandelstam variable $t=(n-p)^2$ in the range
\begin{eqnarray}
%\nonumber
0\leq |t| \leq 0.16~(GeV/c)^2.
\label{range}
\end{eqnarray}

\begin{figure}[t]
%\vskip -4cm
 \epsfxsize=90mm
 \epsfysize=90mm
 \centerline{
 \epsfbox{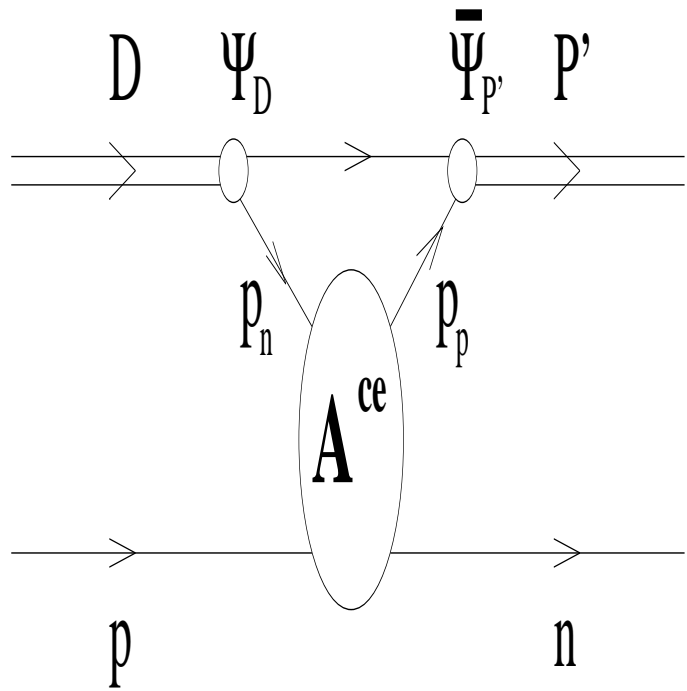}}
 \caption{Graphical representation of the reaction mechanism.}
 \label{pic1}
\end{figure}

In Fig.~\ref{pic1} the assumed reaction mechanism is represented.
The elementary subprocess of the proton-neutron
scattering is denoted by $\CA^{ce}$.
Within the range of $t$ given in (\ref{range})
the subprocess corresponds to the
scattering into the backward hemisphere in the c.m.s. of nucleons
by an elementary charge-exchange process.
Another important kinematical
restriction for this reaction is the low excitation energy of the
final pp pair. This reaction type was experimentally investigated in
Saclay for the following range of invariant masses
of the pp pair:
\begin{eqnarray}
%\nonumber
s_p=P'^2, \quad 4\,m^2\leq s_p \leq (2\,m+8~MeV)^2,
\end{eqnarray}
up to the excitation energy of $E_x = 8 MeV$
(here $P'$ is the total 4-momentum of the pair).
For such low relative momenta it seems to be
very probable that the relative orbital momentum of the pair vanishes.
Hence, due to the Pauli principle, it must be in a spin-singlet state.
One of our basic assumptions is therefore that the final pp pair is
in a $^1S_0$ state with total angular momentum $J=0$.

In Fig.~\ref{pic1}, $\Psi_D$ is the BS amplitude of the deuteron
and ${\bar \Psi}_{P'}$ denotes the conjugate amplitude of the $pp$ pair;
$\CA^{ce}$ is the amplitude of the
elementary charge-exchange process.
In the present calculations we
consider the case of a polarized deuteron with a density
matrix being diagonal with respect to the $z$ axis
(here the $z$ axis is parallel to the momentum $\bp$
of the initial proton).
Assuming a $^1S_0$ final state
of the pp  pair the differential cross
section can be written as
\begin{eqnarray}
%\nonumber
\frac{d^2\sigma}{dt\,ds_p}\,=\,\frac{1}{(8\pi)^3\lambda}
\sqrt{1-\frac{4m^2}{s_p}}\,|M_{fi}|^2.\label{cross}
\end{eqnarray}
It has such a simple form due to the axial symmetry along
the $z$ axis appearing in this case.
The experimental data from Saclay are binned into the following intervals of
excitation energy:
\begin{eqnarray}
I:   && \quad 0\leq E_x \leq 1 MeV,    \nonumber\\
II:  && \quad 1~MeV\leq E_x \leq 4 MeV,\nonumber \\
III: && \quad 4~MeV\leq E_x \leq 8~MeV.\nonumber
\end{eqnarray}
To compare the cross section (\ref{cross}) with the experiment data
it is necessary to integrate over the invariant mass of the pair in the
regions $I$, $II$ and $III$ according to
\begin{eqnarray}
\left(\frac{d\sigma}{dt}\right)_{I,II,III}\,=\,
\frac{1}{(8\pi)^3\lambda}\,
\int\limits_{I,II,III}\,ds_p\,\sqrt{1-\frac{4m^2}{s_p}}\,|M_{fi}|^2.
\end{eqnarray}
The matrix element of the reaction within the BS formalism has the form
\begin{eqnarray}
M_{fi} = \sum\limits_{ss'}\frac{1}{(2m)^2}\int d^4k\,
f_{r's',sr}(s_{pn},t)
{\bar u}^{s}(p_n)\,{\Psi}_D^M(k) \,
[ \frac12 \hat{ D } - {\hat k} + m] \,
{\bar\Psi}_{P'}(k-\frac{q}{2})\,u^{s'}(p_p).
\label{mat}
\end{eqnarray}
The BS amplitude of deuteron $\Psi_D^M$ was obtained as a numerical
solution of the corresponding BS equation with realistic kernel
with 6 one-meson exchanges \cite{umnikov}.
To calculate ${\bar\Psi}_{P'}$ we use the plane-wave approximation
\cite{dorkin} thus disregarding for the moment being
the effects of final state interaction (FSI).
The most
important part of the diagram in Fig.~1, the charge-exchange amplitude
$\CA^{ce}$, is incorporated in the matrix element (\ref{mat})
by the on-shell amplitudes $f_{r's',sr}$ and the Dirac spinors.
In doing so
the off-shell effects are neglected and the elementary subprocess is
considered as real process with on-shell particles. Actually in
the plane-wave approximation only one particle is off-shell, namely
the initial neutron.

In our numerical calculations we use the
helicity amplitudes $H_{\lambda'\nu',\nu\lambda}$ of pn scattering,
resulting from both the Nijmegen partial wave analysis
\cite{nijmeg} and the well-known results of SAID \cite{said}.
In the amplitudes
$f_{r's',sr}$ in (\ref{mat}) the spin indices are not helicities, but
the spin projections on the $z$ axis of the laboratory frame. Thus, to
transform from these projections to helicities,
one has to perform three spin rotations by
\begin{eqnarray}
\nonumber
f_{r's',sr}(s_{pn},t) = \sum\limits_{\sigma\rho'\sigma'}
\CD^{\frac{1}{2}*}_{s\sigma}(p_n)\,\CD^{\frac{1}{2}}_{r'\rho'}(n)\,
\CD^{\frac{1}{2}}_{s'\sigma'}(p_p)\,f^h_{\rho'\sigma',\sigma
r}(s_{pn},t).
\end{eqnarray}
Another important remark concerns the required amplitudes $f_{r's',sr}$
which
are neither in the c.m.s. nor in the laboratory system of the
$pn\to np$ process,
but in some general reference frame defined by the kinematics
of the total process. So, one has to perform a Lorentz boost, and as
a result one gets four additional Wick helicity rotations
\begin{eqnarray}
\nonumber
f^h_{\rho'\sigma',\sigma r}(s_{pn},t) = \sum\limits_{\lambda'\nu'\lambda\nu}
\CD^{\frac{1}{2}}_{\lambda r}(p^w)\,\CD^{\frac{1}{2}}_{\nu\sigma}(p_n^w)\,
\CD^{\frac{1}{2}*}_{\nu'\sigma'}(p_p^w)\,
\CD^{\frac{1}{2}*}_{\lambda'\rho'}(n^w)\,H_{\lambda'\nu',\nu\lambda}(s_{pn},t).
\end{eqnarray}

\begin{figure}[ht]
\vskip 4.5cm
 \epsfxsize=110mm
 \epsfysize=110mm
 \centerline{
 \epsfbox{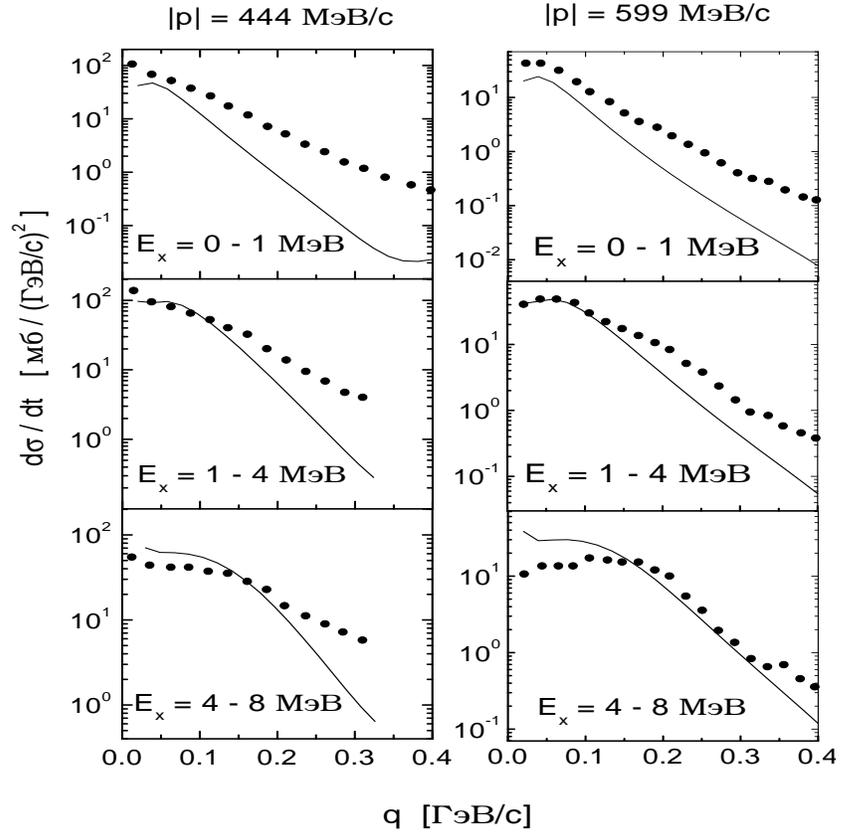}}
 \caption{Differential cross section as a function of $|\bq|$
for two initial momenta
of incoming protons. Solid lines: our calculations, dots: data
from \cite{kox}.}
 \label{pic2}
\end{figure}

\begin{figure}[ht]
\vskip 3cm
 \epsfxsize=110mm
 \epsfysize=110mm
 \centerline{
 \epsfbox{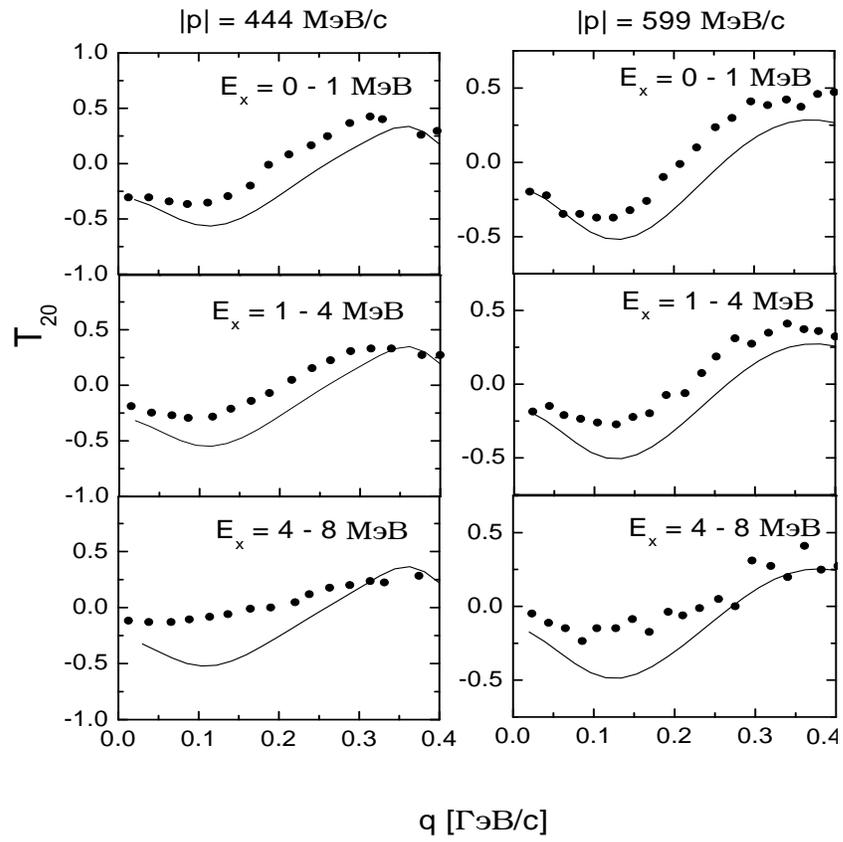}}
 \caption{As in Fig.~2 but for tensor analyzing power $T_{20}$.}
 \label{pic3}
\end{figure}

\begin{figure}[ht]
\vskip 5cm
 \epsfxsize=80mm
 \epsfysize=80mm
\centerline{
\epsfbox{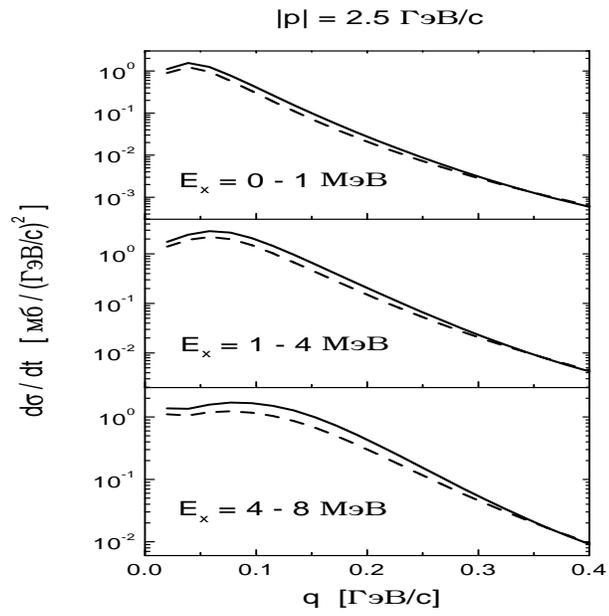}
 }
 \caption{Cross sections as a function of $|\bq|$ for $|\bp|=2.5~GeV/c$.
Solid lines: SAID parameterization, dashed lines:
parameterization from \protect\cite{diu}.}
 \label{pic4}
\end{figure}

\begin{figure}[ht]
\vskip 3cm
 \epsfxsize=80mm
 \epsfysize=80mm
 \centerline{
\epsfbox{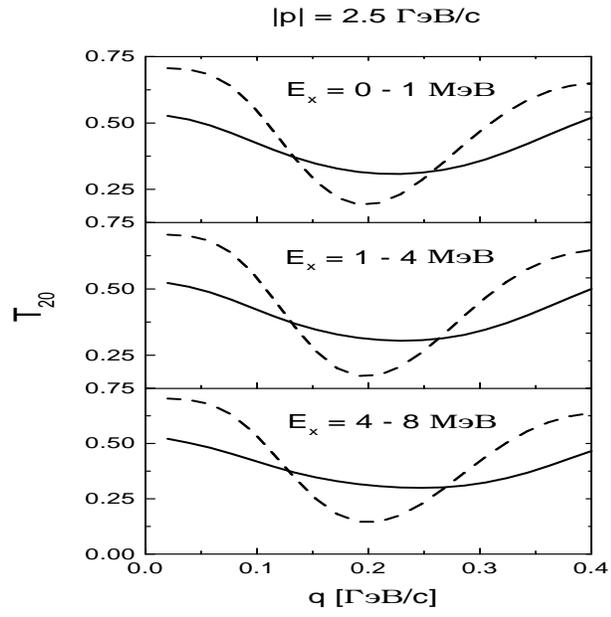}
 }
 \caption{As in Fig.~4 but for the tensor analyzing power $T_{20}$.}
 \label{pic5}
\end{figure}

The results of our numerical calculations are presented in
Figs.~\ref{pic2} - \ref{pic5}. As mentioned above, the
following sets of experimental data from Saclay are at disposal:
the cross section and the tensor analyzing powers were measured for two
values of initial energy (corresponding to the momenta of initial
protons $|\bp|=444$ and $599 \, MeV/c$), and the experimental data were
binned into three intervals of the excitation energies of the
pp pair. As seen in Fig.~\ref{pic2}, the cross section
for the lowest excitation energies is underestimated even for small
values of $|\bq|$ ($|\bq|$  is the three momentum transfer in the
 system where the initial proton is at rest),
while at the higher energies there is a reasonable
agreement with the data. Therefore, the conclusion emerges
that the FSI is not negligible, especially for low values of the
excitation energy. But for higher values of $E_x$ another problem
arises
with the tensor analyzing power $T_{20}$, as can be seen in
Fig.~\ref{pic3}. It is
obvious that a better agreement with the experimental data is obtained for
the lowest values of $E_x$. The discrepancies for higher energies
become greater, although the qualitative agreement with the experiment
still holds. It seems that higher partial waves are to be taken
into account, in particular, the triplet state of the pp pair,
at least when neglecting the FSI.

Now we are in the position to conclude that our model of the reaction
$pD\to n(pp)$ works qualitatively rather well, and
this enforces us to apply it for the case of
relativistic initial energies. The predictions of our model for a
momentum of the initial proton $|\bp|=2.5~GeV/c$ as
available at COSY are presented in Figs.~\ref{pic4} - \ref{pic5}.
The solid lines correspond to the more realistic results obtained
within the SAID parameterization of the elementary charge-exchange
amplitude, while the
dashed lines rely on the parameterization from \cite{diu}.
Our main conclusion is that the
cross sections and tensor analyzing powers $T_{20}$ are large enough
and hence-fore at COSY energies the deuteron charge-exchange
reaction can be used as the basic reaction for a deuteron
tensor polarimeter.

\section{Acknowledgments}
This work was performed in parts during the visits of S.S.S and L.P.K.
in the Forschungszentrum Rossendorf, Institute of Nuclear and Hadron Physics.
We thank for the support by the program "Heisenberg-Landau" of JINR-FRG
collaboration and the grants 06DR921 and WTZ RUS 98/678.

\newpage

\newpage

\newpage

\newpage

\end{document}